\documentclass[aps,prb,twocolumn,showpacs,superscriptaddress,floatfix,10pt]{revtex4-1}
\usepackage{amsmath,amssymb,amsfonts,stmaryrd,wasysym,graphicx,multirow,color,textcomp,subfigure}
\usepackage{url}
\usepackage[colorlinks=true,citecolor=blue,urlcolor=blue]{hyperref}
\usepackage{romannum}

\def\be{\begin{equation}}
\def\ee{\end{equation}}
\def\bea{\begin{eqnarray}}
\def\eea{\end{eqnarray}}
\def\d3{D\Romannum{3}}

\begin{document}
\title{Emergent Localization in Dodecagonal Bilayer Quasicrystals}
\author{Moon Jip Park}
\author{Hee Seung Kim}
\author{SungBin Lee}
\affiliation{Department of Physics, KAIST, Daejeon 34141, Republic of Korea}
\begin{abstract}
A new type of long-range ordering in the absence of translational symmetry gives rise to drastic revolution of our common knowledge in condensed matter physics. Quasicrystal, as such unconventional system, became a plethora to test our insights and to find exotic states of matter. In particular, electronic properties in quasicrystal have gotten lots of attention along with their experimental realization and controllability in twisted bilayer systems. In this work, we study how quasicrystalline order in bilayer systems can induce unique localization of electrons without any extrinsic disorders. We focus on dodecagonal quasicrystal that has been demonstrated in twisted bilayer graphene system in recent experiments. In the presence of small gap, we show the localization generically occurs due to non-periodic nature of quasicrystal, which is evidenced by the inverse participation ratio and the energy level statistics. We understand the origin of such localization by approximating the dodecagonal quasicrystals as an impurity scattering problem. 
\end{abstract}
\maketitle


The theoretical core of solid state physics lies on the crystalline order of atoms. As long as the translational symmetry is present, the eigenstates can be expressed as  modified plane waves having a definite momentum, according to the Bloch theorem\cite{Bloch1929}. On the other hand, quasicrystal, a new type of long range order, lacks the periodicity of crystals thus well-established electronic band structure analysis is not applicable\cite{PhysRevLett.53.1951,doi:10.1146/annurev.pc.42.100191.003345,Steurer2018}. One of the common features in quasicrystals is the low conductivity so electrons are fairly localized in space\cite{Pierce737,Levi1541,PhysRevB.54.12793,Jeon2017,PhysRevB.55.2890}. In 1D and 2D cases, it has been extensively studied that  delocalized, localized and intermediate (often termed as `critical') electronic states may coexist as a unique feature in quasicrystals\cite{PhysRevB.34.2207,PhysRevB.55.2890,Villa:06,PhysRevB.35.1020,PhysRevB.43.8879,PhysRevLett.56.2740,Rotenberg2000,PhysRevB.33.2184,PhysRevB.39.9904,PhysRevB.61.3377}.  However, the coherent understanding in the origins of such electronic properties of quasicrystals are still lacking. This is mainly because there is no known systematic way of studying electronic states of quasicrystalline structures. 

In this regard, recent experimental demonstrations of the dodecagonal quasicrystalline twisted bilayer graphene(TBG) has introduced an interesting platform to study the quasicrystalline structure in a controllable way\cite{Ahneaar8412,Yao6928}. When the graphene layers are not interacting with each other, each layer possesses the full lattice translational symmetry. The low energy physics is governed by the two separate massless Dirac Hamiltonian. Even if a weak interlayer interaction is turned on, the effective translational symmetric Hamiltonian can be written down using the perturbative approach\cite{1367-2630-17-1-015014}, which instead realizes infinite numbers of replica Dirac cones\cite{Ahneaar8412,Yao6928}. However, if the interlayer coupling strength is strong enough to generate a significant correlation of the bilayer, such a perturbation approach is not any more valid, and the TBGs fully lost the translational symmetry. In this strongly correlated regime, we expect the full dodecagonal quasicrystalline order such that the behavior of the electrons cannot be described by the Bloch wave function. The understanding of the electronic states in this regime is now completely missing.

%

\begin{figure}[t!]
	\centering\includegraphics[width=0.5\textwidth]{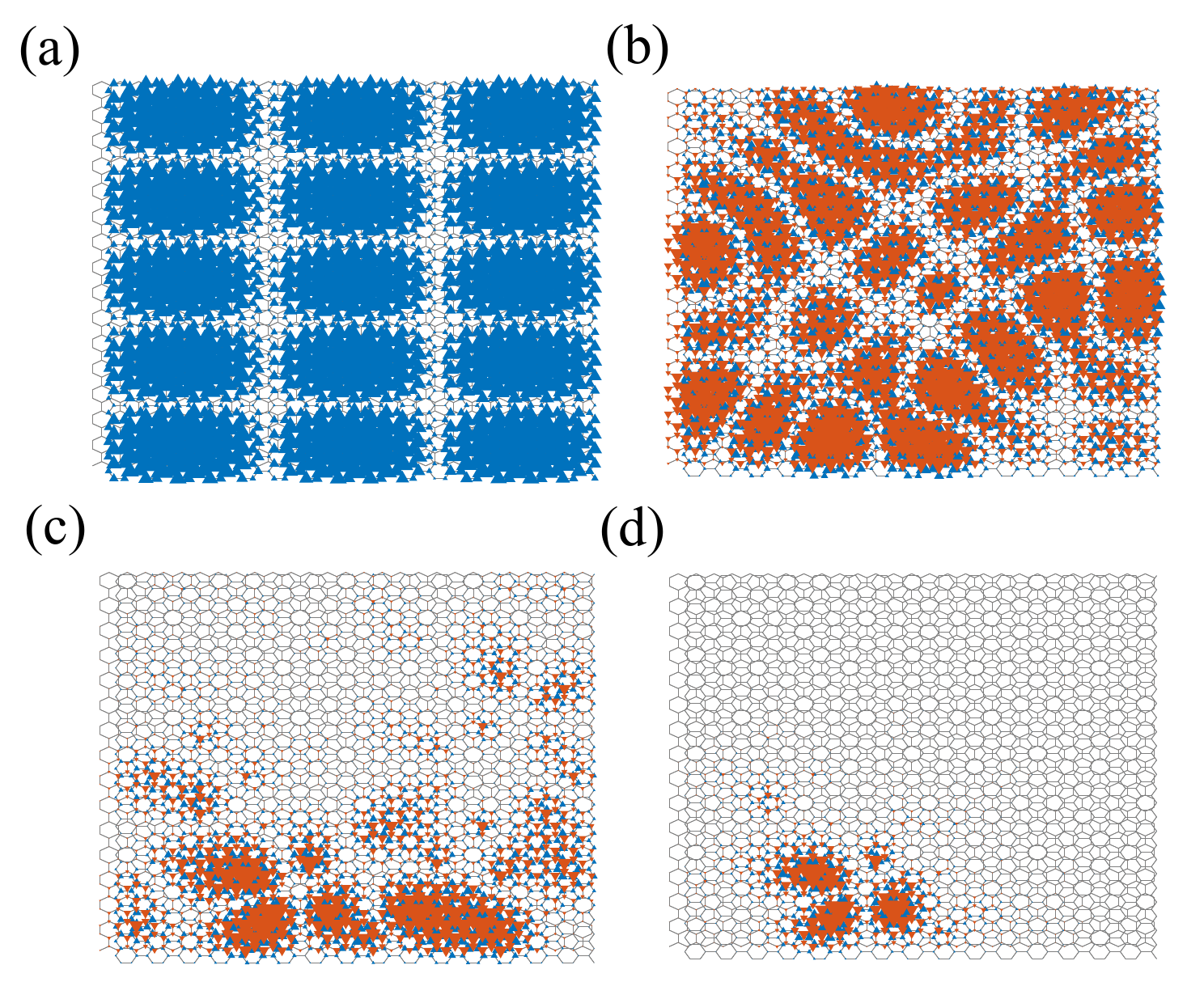}
	\caption{Illustration of a single eigenstate in dodecagonal quasicrystal as a function of the interlayer coupling strength, $V_{int}/t=0,1,4,6$ for (a)-(d) respectively. We set $m_{top}=m_{bot}=2t$. Blue (red) triangles represent the distribution of wave function magnitudes in top (bottom) layer respectively. With increasing interlayer coupling strength, the eigenstates between top and bottom layers begin to correlate with each other and become localized for large $V_{int}$.
	}\label{fig:1} 
\end{figure}

In this work, we report how electronic properties in dodecagonal quasicrystals can be controlled and show the emergence of electronic localization as the intrinsic effect due to quasiperiodicity.
 Our model consists of a honeycomb bilayer that has the relative twist angle, $\theta=30^\circ$. When the sublattice symmetry breaking mass term is considered, the eigenstates near the band edge starts to be localized. This behavior directly contrasts with TBGs at commensurate angles, where the states are fully delocalized over the extended unit cell even with the arbitrary strength of interlayer coupling\cite{PhysRevLett.99.256802,Bistritzer12233,PhysRevB.83.045425,PhysRevB.86.155449,PhysRevB.85.195458,PhysRevB.87.205404,doi:10.7566/JPSJ.84.121001,PhysRevB.86.125413,PhysRevX.8.031087}.  To understand the origin of the localization, we show that we can approximate our bilayer model as a monolayer impurity scattering model. We solidify our results using numerical diagonalization of the tight-binding model. This localization behavior is further quantitatively analyzed by the calculation of the inverse participation ratio(IPR)\cite{Wegner1980,PhysRevLett.84.3690,PhysRevLett.69.1093} and the energy level statistics\cite{PhysRevB.96.214201,PhysRevB.75.155111,PhysRevB.82.174411}. Especially, we show the energy-dependent transitions from GOE to Poisson distribution, which imply the existence of the mobility edge\cite{PhysRevB.22.5823}.

We begin our discussion by writing down the Hamiltonian, $H_{tot}$, that describes the interacting honeycomb bilayer system, which is given as\cite{Bistritzer12233,Ahneaar8412},
\begin{gather}
H_{tot}\!=\!
\sum_{i,j}\left(
\begin{array}{cc}
c_{T,i}^\dagger \\
c_{B,i}^\dagger \\
\end{array}
\right)^T \! \! \!
\left(
\begin{array}{cc}
(h_{top})_{i,j} \! & (g_{\text{int}})_{i,j} \! \\
(g_{\text{int}}^\dagger)_{i,j} \! & (h_{bot})_{i,j} \! \\
\end{array}
\right) \!
\left(
\begin{array}{cc}
c_{T,j} \\
c_{B,j} \\
\end{array}
\right),
\label{Eq:Ham1}
\end{gather}
where $h_{top}$ and $h_{bot}$ are the matrices of the Hamiltonians that describe the top and bottom monolayer honeycomb lattice with the nearest neighbor hopping, $t$, and the sublattice symmetry breaking mass term, $m_{top}$ and $m_{bot}$ respectively. $c_{T,i}(c_{B,i}^\dagger)$ is the fermion annihilation(creation) operator at site $i$ on the top (bottom) layer. 
The sites $i$ on the top and the bottom layer are associated with the real space coordinate, $\vec{R}_{T,i}$ and $\vec{R}_{B,i}$. These vectors are related by the two-dimensional rotational matrix, $M(\theta)$ for rotation angle $\theta$ along $z$ direction and the distance $\vec{r}_z$ between top and bottom layers, thus $\vec{R}_{T,i}=M(\theta)\vec{R}_{B,i} + \vec{r}_z$. The interlayer coupling, $g_{\text{int},i,j}\equiv G_{\text{int}}(|\vec{R}_{T,i}-\vec{R}_{B,j}|)$, is given by the transfer integral between the two $p_z$ orbitals of the two atoms. Using the Slater-Koster formula, we can write the transfer integral between the layer as a simple exponentially decaying function,
\begin{gather}
G_{\text{int}}(|R|)=V_{int}e^{-|R|/\delta_0}.
\end{gather}
The parameter $V_{int}$ for transfer integral can be estimated by both first principle calculation and experimental observation, which is yet controversy\cite{PhysRevB.86.155449,Ahneaar8412}. 
From now on, we set particular values for decay length $\delta_0=0.184a$\cite{doi:10.1021/nl902948m,PhysRevB.69.075402} and $\vec{r}_z\!=\!0$ where $a$ is the lattice constant, and analyze the electronic structure as a function of $V_{int}$. We emphasize that our analysis is still valid for different values of $\delta_0$ and finite $\vec{r}_z$ by rescaling the overall energy scale of $V_{int}$.

For certain discrete sets of the twist angles, the TBGs recover the periodic translational symmetry by producing Moire patterns\cite{PhysRevLett.99.256802}. These angles are given by $\cos(\theta_{p,q})=\frac{3p^2+3pq+q^2/2}{3p^2+3pq+q^2}$
where $p$ and $q$ are arbitrary coprime integers\cite{PhysRevB.87.245403,PhysRevLett.99.256802}. In this case, the unit cell of the TBG is enlarged by $L=a_0/(2\sin(\theta_{p,q}/2))$. Unlike the commensurate angles, the twisted angle at $\theta=30^\circ$ realizes the dodecagonal quasicrystalline order and the system completely loses the translational symmetry. Nevertheless, the rotational symmetries are still intact. On symmetry grounds, the dodecagonal quasicrystal possesses $C_{12}$ rotation symmetry in the massless limit($m_{top}=m_{bot}=0$) otherwise only $C_3$ symmetry is preserved.

\begin{figure}[t!]
	\centering\includegraphics[width=0.5\textwidth]{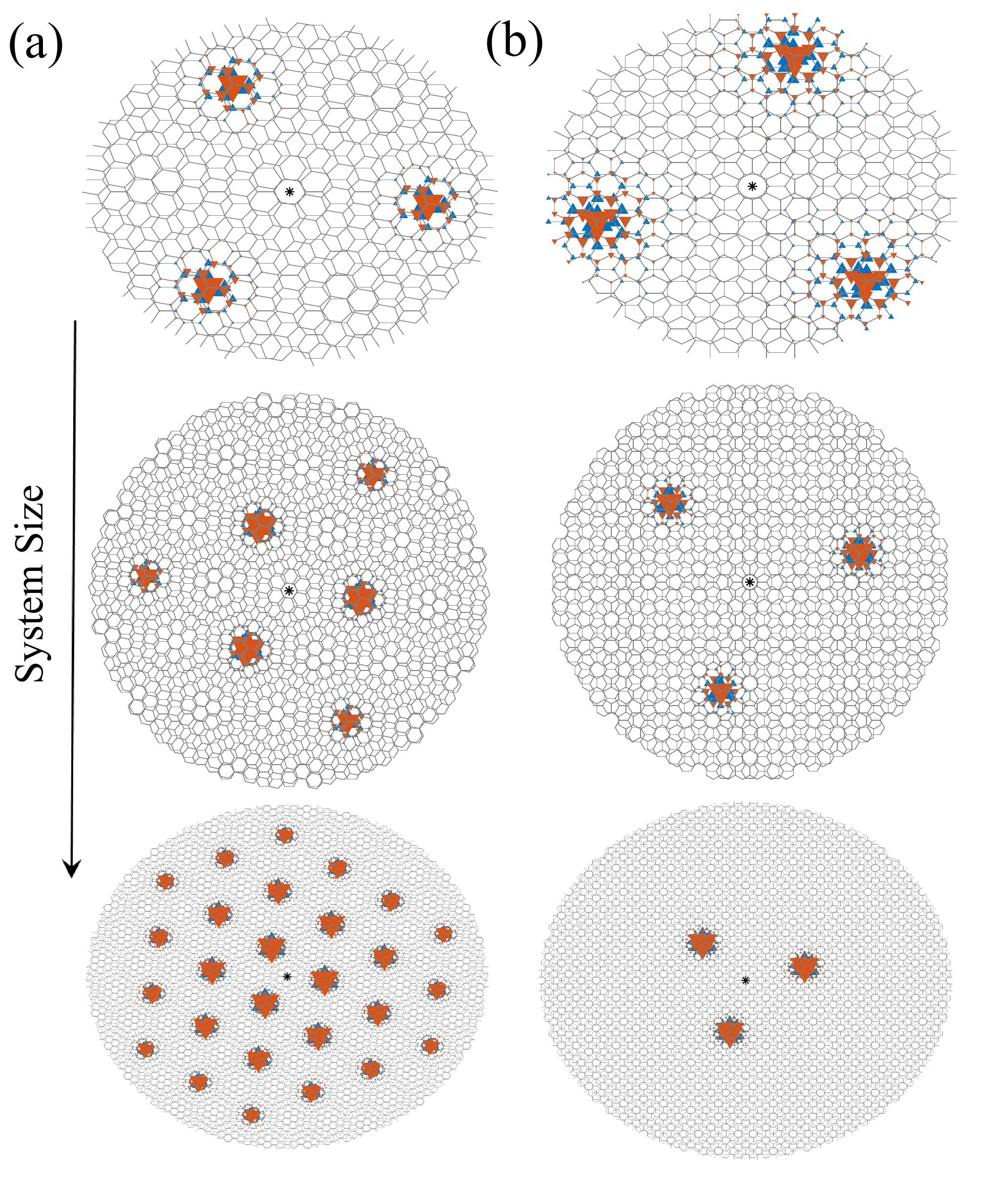}
	\caption{The comparison of the localized states at (a) commensurate angle($p=7,q=3$) and (b) the dodecagonal quasicrystal over various system sizes. The states at the commensurate angle can be only localized within the extended unit cells, while the states in the dodecagonal quasicrystal are fully localized. $m_{top}=m_{bot}=2t$ and $V_{int}=6t$ are used. }\label{fig:spectralbottom} 
\end{figure}
\begin{figure}[t!]
	\centering\includegraphics[width=0.5\textwidth]{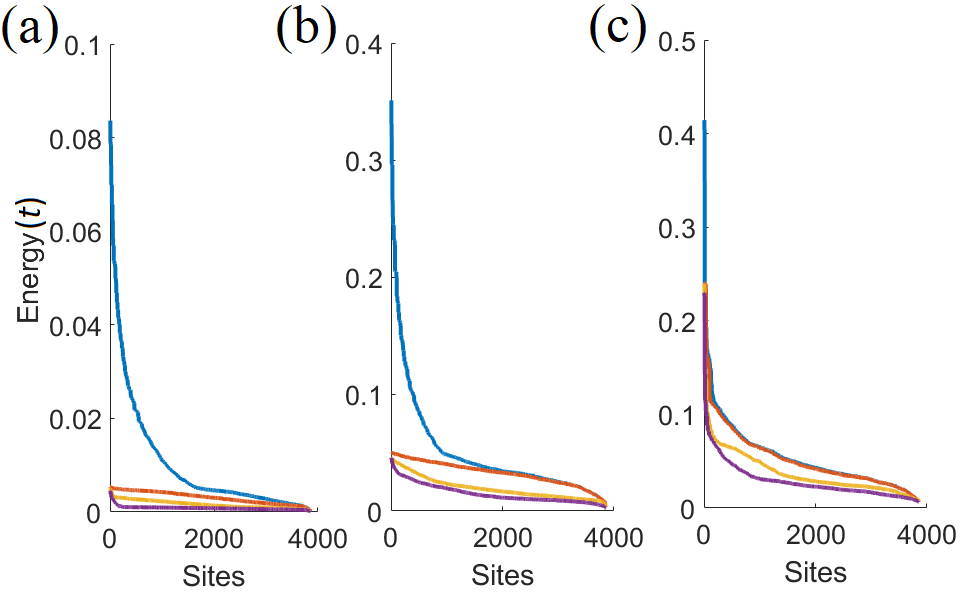}
	\caption{Distribution of $\Sigma_{imp}(E=0)$ in descending order with (a)  $m_{bot}=10t$ (b) $m_{bot}=t$ and (c) $m_{bot}=0.1t$. Blue line represents the on-site component of $\Sigma_{imp}$. Other colors represent the off-diagonal components.}\label{fig:dist} 
\end{figure}

In order to analyze the electronic spectrum, we perform exact diagonalization of the tight-binding Hamiltonian in Eq.\eqref{Eq:Ham1} with a given system size up to 10000 sites. 
Fig. \ref{fig:1} shows the density profile of a single eigenstate near the band edge as a function of the interlayer coupling, $V_{int}$ when $m_{top}=m_{bot}=2t$. Blue (red)  triangles represent the distribution of wave function magnitudes in top (bottom) layer respectively.  In the non-interacting case, $V_{int}=0$, Fig. \ref{fig:1} (a) shows a fully extended wave packet, as each layer is described by the Dirac Hamiltonian. In this case, we only show the density profile of the eigenstate near band edge for the top layer (blue triangles) which is identical to the one for the bottom layer. When the interlayer coupling strength increases, the top and bottom layers begin to correlate each other. As shown in Fig. \ref{fig:1} (b) and (c), the corresponding density profiles of the eigenstates are not a simple wave packet since the interlayer coupling breaks the translational symmetry. Further increasing the interlayer coupling strength, we discover the emergence of the localized eigenstates as shown in Fig. \ref{fig:1} (d).

Such localization behavior induced in the dodecagonal bilayer quasicrystal is qualitatively distinct from the localizations observed in bilayer Moire patterns realized with the commensurate angles between layers\cite{PhysRevB.86.125413,PhysRevB.87.245403,PhysRevB.90.155451}. In the bilayer Moire patterns, the wave functions are localized only within the extended unit cells, but form extended state at the longer distance than the unit cells which can be explained with enlarge unit cell. However, in the dodecagonal bilayer quasicrystal, the wave functions are fully localized and do not extend in longer distances. This qualitative difference can be explicitly checked by comparing the eigenstates for the bilayer Moire patterns and the dodecagonal quasicrystal. Fig. \ref{fig:spectralbottom} (a) and (b) compare the eigenstates near the zero energy for particular Moire pattern with the commensurate angle $\theta=\cos^{-1}(143/146)$ ($p=7,q=3$) and the dodecagonal quasicrystal with $\theta=30^\circ$. For small system size, the eigenstates in both cases are well-localized and only maintain $C_3$ symmetry. However, as the system size increases larger than the size of  extended unit cells for Moire pattern we take into account here, the additional Wannier centers appear in the case for the commensurate angle forming an extended state. (See Fig.\ref{fig:spectralbottom} (a).) Unlike the commensurate angle, we observe the wave function is completely localized even for very large system size in the case of  dodecagonal bilayer quasicrystal as shown in Fig.\ref{fig:spectralbottom} (b). In this case, we only see the existence of three-fold symmetry by averaging out all the degenerate eigenstates.

In order to understand such emergent localization, one can consider the extreme limit where the mass of bottom layer is very large i.e., $m_{bot}\gg t$. 
The eigenstates of $H_{tot}$ in Eq.\eqref{Eq:Ham1} satisfy the following equation,
\begin{gather}
h_{top} u_n+g_{int} v_n = E_n u_n,
\nonumber
\\
g_{int}^\dagger u_n +h_{bot} v_n = E_n v_n.
\label{Eq:3}
\end{gather}
Here, $u_n$ and $v_n$ are the $n$-th wave vectors of the top and bottom layers respectively. $E_n$ is the corresponding $n$-th energy. One can rewrite Eq.\eqref{Eq:3} in terms of the top layer wave function, $u$, by introducing the self-energy term,
\begin{gather}
{(h_{top})}_{ij} u_{n,j}+\Sigma_{imp}(E_n)_{ij} u_{n,j} = E_n u_{n,i},
\label{Eq:Sig}
\end{gather}
where the self-energy term is explicitly given as, $\Sigma_{imp}(E_n)=g_{int}\frac{1}{E_n-h_{bot}}g_{int}^\dagger$.
Now, the above equation is equivalent to the single layer honeycomb lattice Hamiltonian with a quasi-periodic self-energy, $g_{int}\frac{1}{E_n-h_{bot}}g_{int}^\dagger$, which follows the translational symmetry of the bottom layer. Eq. \eqref{Eq:Sig} should be solved self-consistently in a sense that the Hamiltonian is again the function of the energy, $E_n$. Rather than exactly solving this problem, we now focus on the energy near the Dirac points, $E_n\approx0$. 

In the limit of $m_{bot}\gg t$, the eigenstates of $h_{bot}$ near the zero energy are given as the evanescent wave that decays as $\sim e^{-m_{bot} |r|/t}$. As a result, the most dominant contribution of $\Sigma_{imp}(E_n)$ comes from the diagonal components when $\delta_0$ is small enough. Thus, one can approximate $\Sigma_{imp}(E_n)$ as effective on-site impurity scattering potentials. The validity of this approximation can be also numerically checked by calculating the distribution of $\Sigma_{imp}(E_n=0)_{ij}$. Fig. \ref{fig:dist} shows the distribution of $\Sigma_{imp}(E_n=0)_{ij}$ for various values of $m_{bot}$. Fig. \ref{fig:dist} (a) confirms that the on-site components are the dominant contributions of the self-energy in the large mass limit. Even when $m_{bot}\sim t$, Fig. \ref{fig:dist} (b) shows that the on-site components are still dominant. Under this approximation, one can regard the bilayer Hamiltonian in Eq.\eqref{Eq:Ham1} as the problem of monolayer with quasi-periodic impurity potentials,
\begin{gather}
{h_{top}}_{ij} u_{j,n}+ V_{imp}\sum_{\vec{R'_j}\in bottom} e^{-2|\vec{R_i}-\vec{R'_j}|/\delta_0}u_{i,n} = E_n u_{i,n}.
\end{gather}
Here, the impurities are located at the position of atoms in the bottom layer. This type of the impurity scattering has been widely studied when the impurities are randomly distributed. It is well-known that the localization transition occurs if the impurity scattering is short ranged or strong enough to induce the intervalley scattering\cite{PhysRevLett.97.236802,PhysRevB.76.214204,PhysRevLett.102.106401,doi:10.1143/JPSJ.67.1704,PhysRevB.77.115109,PhysRevB.93.235426}. Therefore, we expect the localization of the subgap states $(E_n \ll m_{bot})$.
Unlike the case $m_{bot} \gg t$, when $m_{bot} \lesssim t$, the self-energy $\Sigma_{imp}$ acquires effective long-range hopping terms. This largely enhances non-local hopping processes in Eq. \eqref{Eq:Sig} through the interlayer coupling. This scenario can be also numerically checked. Fig. \ref{fig:dist} (c) shows that the off-diagonal terms of $\Sigma_{imp}(E=0)$ become comparable to the on-site term. In this case, the previous impurity model approximation is not any more valid. We expect the disappearance of the fully localized states.

\begin{figure}[t!]
	\centering\includegraphics[width=0.5\textwidth]{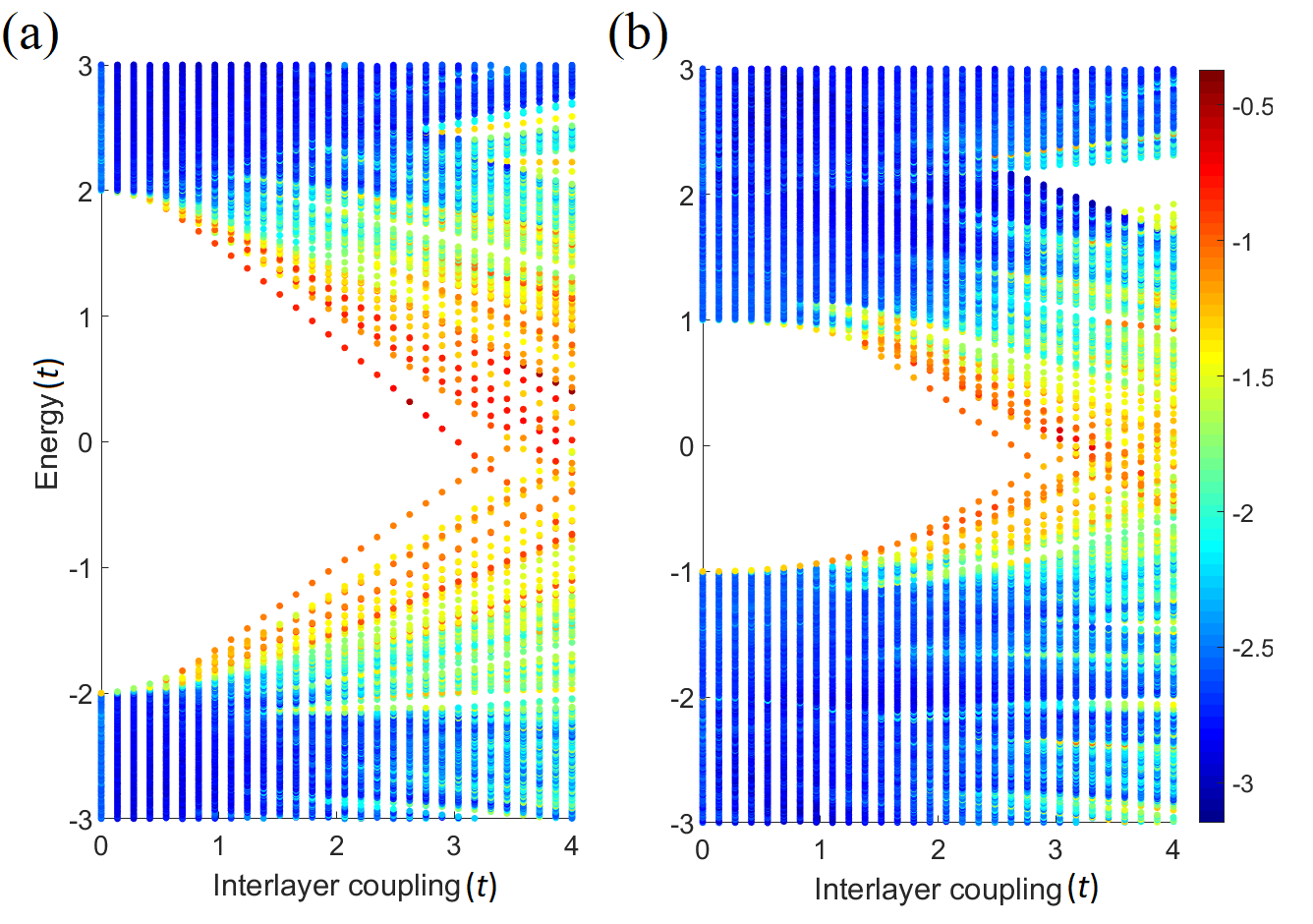}
	\caption{Evolution of the IPR and the energy spectrum as a function of the interlayer coupling when (a) $m_{top}=m_{bot}=2t$ and (b) $m_{top}=t,m_{bot}=2t$. The color of each dot represents the log value of the corresponding IPR.}\label{fig:IPR} 
\end{figure}

To confirm the expected behaviors derived from the impurity model approximation, we calculate the IPR of the eigenstates. The IPR quantifies how much wave functions are spatially extended over the system. The IPR is explicitly defined as,
\begin{gather}
\text{IPR}_i=\sum_i |\psi_n(i)|^4/(\sum_i |\psi_n(i)|^2)^2,
\end{gather}
where $\psi_n(i)$ is $n$-th eigenstates derived from the diagonalization, and $i$ represents the site index. The $\text{IPR}$ of an extended state scales as $1/V$, and converges to zero in a large system limit. On the other hand, the $\text{IPR}$ of a localized state remains $O(1)$ constant as $\sum_n |\psi_i(n)|^4$ do not decay as a function of the system size. Fig. \ref{fig:IPR} (a) and (b) shows the calculation of the IPR when the mass terms of the top and bottom layers are symmetric ($m_{top}=m_{bot}=2t$) and asymmetric ($m_{top}=t$, $m_{bot}=2t$) respectively. In the large mass limit, Fig. \ref{fig:IPR} shows the appearance of the subgap states. The corresponding IPR value indicates that these subgap states are localized, while the supragap states are still delocalized. We can understand this behavior by considering the validity of the impurity model approximation. At the subgap energy, the on-site terms are the dominant contribution of $\Sigma_{imp}$. On the other hand, the resonant states exist at the supragap energy. $\Sigma_{imp}$ again become non-local. Therefore, the impurity approximation is invalidated at the supragap energies. This energy-dependent localization behavior resembles the mobility edge in the Anderson model.  

\begin{figure}[t!]
	\centering\includegraphics[width=0.5\textwidth]{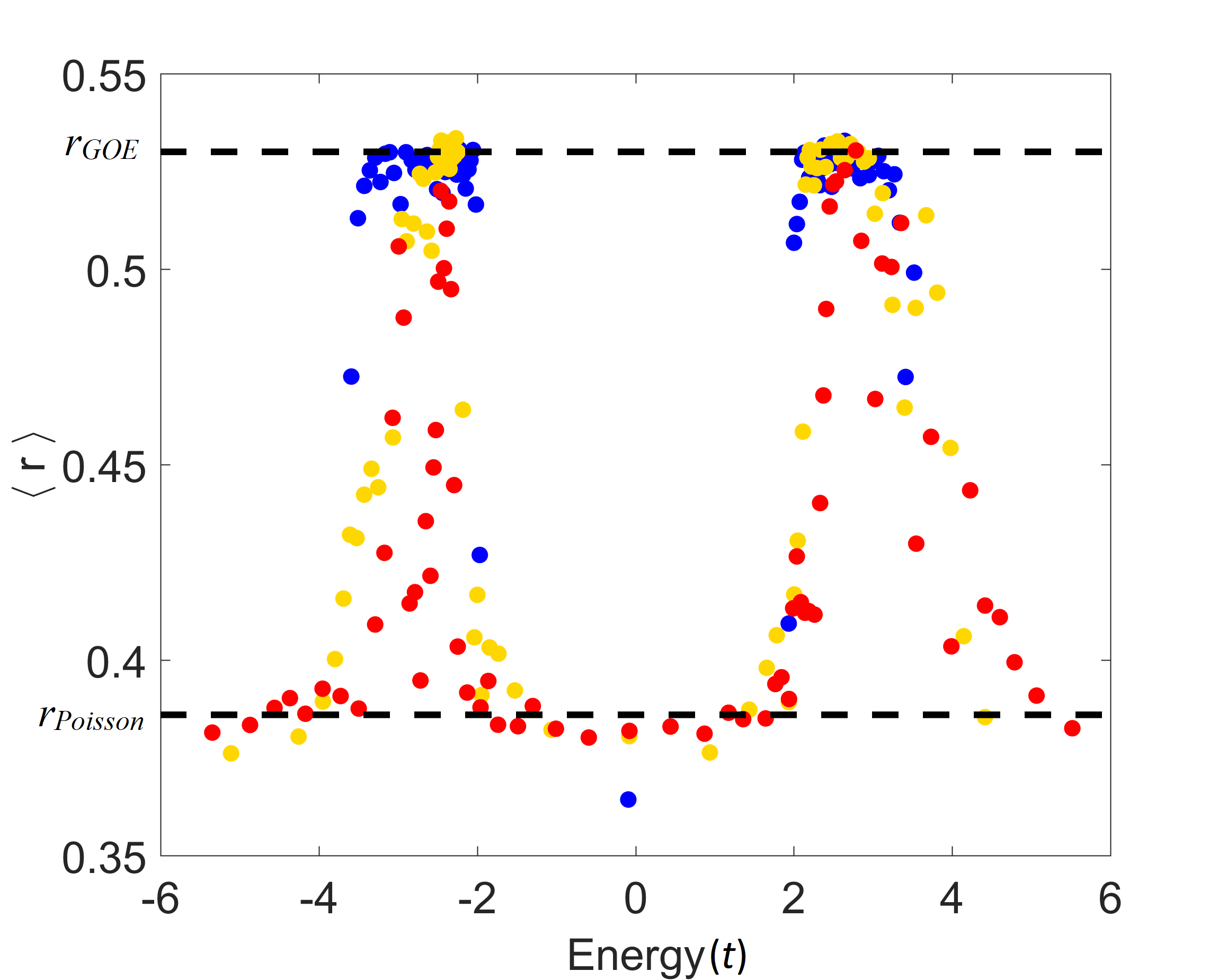}
	\caption{The level spacing ratio as a function of the energy and the interlayer couping strength $V_{int}=t,5t,10t$(blue,yellow,red). $r_{GOE}\approx 0.530$ indicates $GOE$ statistics and $r_{Poisson}\approx0.386$ indicates the Poisson statistics. Black horizontal lines represent $r_{GOE}$ and $r_{Poisson}$ respectively.}\label{fig:levelstat} 
\end{figure}

To solidify the localization behaviors, we can also calculate the level statistics\cite{PhysRevB.96.214201,PhysRevB.75.155111,PhysRevB.82.174411} in addition to the calculation of the IPR. The level spacing ratios are defined as $r_n=min{(\delta_n,\delta_{n+1})}/max{(\delta_n,\delta_{n+1})}$ where $\delta_n$ is the level spacing between $(n+1)$-th and $n$-th energy eigenvalue. When the eigenstates are delocalized, the eigenstates experience the level repulsions. The corresponding level statistics is described by Gaussian orthogonal ensemble(GOE) distribution, and the level spacing ratio converges to $r_{GOE}\approx 0.530$. However, when the states are localized, the distribution of the energy eigenvalues becomes completely random. This scenario is described by Poisson statistics with the corresponding level spacing ratio $r_{Poisson}\approx0.386$. Fig. \ref{fig:levelstat} shows the mean value of the level spacing ratio, $r$, when the symmetric mass terms are included i.e., $m_{top}=m_{bot}=t$. We sample the eigenstates while varying the displacement between the layers\cite{PhysRevB.96.214201}. In the weak interacting limit, we observe that the $r$ of the supragap states converges to $r_{GOE}$. However, as the interaction strength increases, we find that the subgap states converges to $r_{Poisson}\approx 0.386$, indicating the localization behavior. The overall behavior of the level statistics confirms the localization behavior observed from the calculation of the IPR.

In conclusion, we have studied the electronic states of the dodecagonal bilayer quasicrystals constructed by two layers of honeycomb lattice with twisted angle $\theta=30^\circ$. Based on numerical diagonalization and confirmed by calculation of the IPR and level statistics, we have shown that the system possesses localization of electronic states in the presence of finite sublattice symmetry breaking mass term. In addition, the localization behavior can be understood by approximating the bilayer quasicrystal into the monolayer with the impurity disorder. Such localization behavior is purely intrinsic without any extrinsic disorders, and is uniquely developed due to quasiperiodicity of bilayer system. Further studies on the critical behavior of this localization transition would be also interesting topic of future study.

\acknowledgments
We thank Pilkyung Moon for fruitful discussion. M.J.P and S.B.L. are supported by the KAIST startup and National Research Foundation Grant (NRF-2017R1A2B4008097). 

\bibliography{reference}

\end{document}